# Influence of Electron –Diffusion On The Wave Front Shape for Intense Shock Waves in Xenon

**R. Annou**[1] and **B. Ferhat**[2]
[1] Theoretical Physics Laboratory,
[2] Quantum Electronics Laboratory,
Faculty of Physics, USTHB, Algiers (ALGERIA).

**Abstract.** A bifurcation of strong shock waves front for xenon in a shock tube has been noted. We propose a model which explains the effect and gives a variation-law of the front shape with respect to the Mach number $M$ and the unperturbed gas pressure $p_1$.

For particular Mach numbers M, the front of strong shock waves in xenon becomes unstable. (Ryazin 1980; Egorushkin *et al.*, 1990, Annou and Ferhat, 1997). The shape of the front appears to be a triangle as shown in figures (1) and (2) (Ferhat, 1979). Measurements have shown the linear dependence of the triangle area on M and the proportionality coefficient dependence on the unperturbed gas pressure $p_1$. In this letter we propose a model which includes electron – diffusion to the walls of the shock tube. It could be an acceptable first step in understanding and quantizing the effect above discussed.

The extension of the wave front $d$ is determined by the longest relaxation process, which is ionization. Hence,

$$d \approx v t_{ion} = M c_1 t_{ion}, \qquad (1)$$

where $t_{ion}$ is the ionization relaxation time, $c_1$ the sound velocity and v the front velocity. Unfortunately, $t_{ion}$ is not known with precision as it is the case with the vibration relaxation time $t_{vib}$. Moreover, it has been noted experimentally that $t_{ion}$ can be related to $t_{vib}$ through a coefficient S. therefore we take $d$ as a function of $t_{vib}$, i.e.,

$$d = M c_1 S t_{vib}, \qquad (2)$$

with,

$$t_{vib} = \frac{1}{Z\left[1-\exp\left(-\frac{\hbar w}{kT}\right)\right]\exp\left[-3\left(\frac{p^2 w^2 m}{2r^2 kT}\right)^{\frac{1}{3}}\right]} \qquad (3)$$

where $m$ is the reduced mass of the particles in collision, $1/r$ is the intermolecular interaction radius and $w$ is the circular frequency of the intermolecular vibrations. Let us expand $t_{vib}$ in the neighborhood of the infinity, since temperature $T$ is very high. One finds,

$$d \approx \frac{S(g+1)}{Z}\sqrt{\frac{R\hbar w}{k(g-1)}}\left[\left(\frac{k}{\hbar w}\right)^{3/2}T^{3/2} - \left(\frac{k}{\hbar w}\right)^{7/6}\left(\frac{p^2 mw}{2r^2}\right)^{1/3}T^{7/6} - \right.$$

$$\left. -\left(\frac{k}{\hbar w}\right)^{5/6}T^{5/6} - \left[1+3\left(\frac{p^2 mw}{2\hbar r^2}\right)^{1/3}\right]\left(\frac{k}{\hbar w}\right)\sqrt{T}\ldots \right] \quad (4)$$

After having established dependence between $d$ and $T$, it's easy to show that the bifurcation can be caused by a temperature gradient along the tube radius. Indeed, if we take into account the electron diffusion towards the walls, the temperature decreases along the diffusion path. Since electrons are $1/r^2$ interacting particles (Zeldovich and Raizer, 1966). This may be easily understood, since the loss of energy occurs at the walls. Moreover, the temperature is a quadratic function of Mach number $M$, so it decreases along the radius.

$$T \to \frac{2 g T_1 (g-1)}{(g+1)^2} M^2 \qquad (5)$$

where, $T_1$ is the unperturbed gas and $\gamma$ is the specific heat ratio. Let us take then, a constant temperature gradient, with the constant being negative, i.e.,

$$dT/dy = -A \qquad (6)$$

If we assume $A$, density dependent, we show that the critical Mach numbers corresponding to the neutral atoms saturation in the front (see Egorushkin *et al.*) are the same as those corresponding to an optimum temperature gradient. Indeed, if we differentiate with respect to M Eq.(6), we find,

$$\frac{\partial}{\partial M}\left(\frac{\partial T}{\partial y}\right) = -\frac{\partial A}{\partial M} = -\left(\frac{\partial A}{\partial n}\right)\left(\frac{\partial n}{\partial M}\right) \qquad (7)$$

For high temperature, we establish dependence between the front width and the spatial dimension $y$ using Eq.(6),

$$\gamma = \gamma(y) \approx l_1(-Ay + T_2)^{3/2} + l_2(-Ay + T_2)^{7/6} \qquad (8)$$

where, $l_1, l_2, l_3$ are constants defined by the Eq.(4). The front is no longer plane. It may be noticed that $d$ can be evaluated differently, taking also into account radiation. We find,

$$\gamma \propto \frac{l}{M\left(k_T + \frac{16}{3}\sigma l_R T_1^3\right)} \qquad (9)$$

where $l$ is the mean free path, $k_T$ is the thermal conduction coefficient, $\sigma$ is Stefan's constant, $l_R$ is Rossland's mean free path and $T_1$ is the unperturbed gas temperature.

From Eq.(9), we note that for high M, the front grows thinner.

Let us evaluate now, the variation of $d$,

$$\Delta d \propto \left[\left(\frac{k}{\hbar \omega}\right)^{3/2}\sqrt{T} - \frac{7}{3}\left(\frac{k}{\hbar \omega}\right)^{7/6}\left(\frac{p^2 m \omega}{2\hbar r^2}\right)^{1/3} T^{1/6}\right]\Delta T \qquad (10)$$

By using equation (10), one finds that the front grows thinner below a temperature $T_0$ and blows up when temperature is greater than $T_0$, which is given by,

$$T_0 = \left(\frac{7}{3}\right)^3 \frac{p^2 m \omega^2}{2 r^2 \hbar} \qquad (11)$$

We find finally using equations (5) and (6) (see, Fig.3),

$$\Delta(mm^2) \underset{\infty}{\sim} 1.2\sqrt{P_1(torr)}M , \qquad (12)$$

where, $2\Delta = Y_0 d = Y_0(d(Y_0) - d(0))$

$p_1$=10 torr

| M\\$\underline{\Delta}$ ($mm^2$) | Experimental | Theoretical | Precision |
|---|---|---|---|
| 18 | 73.3 | 68.3 | 6% |
| 22 | 58.3 | 83.5 | 30% |
| 24.4 | 86.7 | 92.6 | 6.5% |
| 27 | 104.2 | 102.5 | 2% |
| 30 | 116.7 | 113.8 | 3% |

$p_1$=25 torr

| M\\$\underline{\Delta}$ ($mm^2$) | Experimental | Theoretical | Precision |
|---|---|---|---|
| 17 | 66.7 | 102 | 35% |
| 18.1 | 100 | 109 | 9% |
| 18.6 | 103 | 112 | 8% |

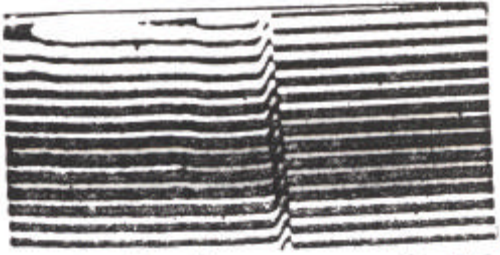

Fig.1: non perturbed plane front. Stainless steel walls. $P_1$=25 torr; M=10.

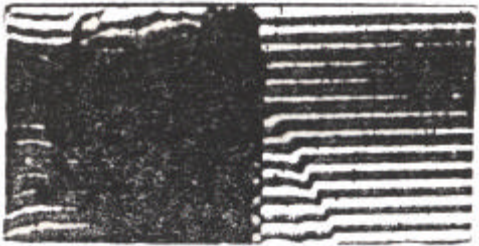

Fig.2: perturbed front. Glass walls. $P_1$=10 torr; M=14

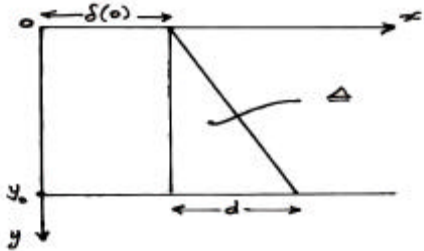

Fig.3: modeling of the front

## CONCLUSION

In shock tubes, and for critical values of the Mach number, the front gets distorted. For some cases, it appears a triangle-like shape for the front. This model explains the fact that the area of the triangle varies linearly with $M$ and the proportionality coefficient is related to $p_1$ as reported in many experiments.